\documentclass[10pt,conference]{IEEEtran}
\IEEEoverridecommandlockouts

\usepackage[latin1]{inputenc}
\usepackage{graphicx, amssymb, amsmath, amsfonts, amsthm}
\usepackage[usenames,dvipsnames]{color}
\usepackage{bbm}

\usepackage[short]{optidef}

\newcommand{\zd}{zeroDiag}
\newcommand{\vect}{vec}

\allowdisplaybreaks
\IEEEoverridecommandlockouts

\begin{document}
\title{Optimizing Reconfigurable Antenna MIMO Systems with Coherent Ising Machines}

\author{\authorblockN{Ioannis Krikidis\authorrefmark{1}, Abhishek Kumar Singh\authorrefmark{2}, and Kyle Jamieson\authorrefmark{2}}
\authorblockA{\authorrefmark{1}Department of Electrical and Computer Engineering, University of Cyprus, Cyprus}
\authorblockA{\authorrefmark{2}Department of Computer Science, Princeton University, NJ, USA \\
E-mail: krikidis@ucy.ac.cy, \{aksingh, kylej\}@princeton.edu}}

\maketitle

\begin{abstract}

Reconfigurable antenna multiple-input multiple-output (MIMO) is a promising technology for upcoming 6G communication systems. In this paper, we deal with the problem of configuration selection for reconfigurable antenna MIMO by leveraging Coherent Ising Machines (CIMs). By adopting the CIM as a heuristic solver for the Ising problem, the optimal antenna configuration that maximizes the received signal-to-noise ratio is investigated. A mathematical framework that converts the selection problem into a CIM-compatible unconstrained quadratic formulation is presented. Numerical studies show that the proposed CIM-based design outperforms classical counterparts and achieves near-optimal performance (similar to exponentially complex exhaustive searching) while ensuring polynomial complexity.   
\end{abstract}


\section{Introduction}

The evolution to 6G communication systems requires breakthrough physical-layer and networking technologies that can support extremely high engineering requirements. Technologies such as reconfigurable intelligent surfaces, Terahertz communications, semantic communications, fluid antenna systems, radiated near-field communications, digital twins \textit{etc.} are just some examples of current research activities to scale up current infrastructure towards 6G \cite{TAT}. All these new communication paradigms significantly increase computation overhead and demand computing resources with extremely high capabilities. However, classical (silicon) computing architectures cannot be further advanced due to transistors reaching their atomic limits~\cite{itrs}. Quantum computing is a promising tool to overcome this computing bottleneck and provide an appropriate computing platform to wireless technologies \cite{KIM}. The application of physics-inspired quantum computing architectures and algorithms in wireless communication systems is a new research area of paramount importance~\cite{KIM,ABH1,ABH2,JAM,JAM2}. 

Quantum computing is built on the fundamental concepts of superposition and entanglement and mainly refers to two basic models \textit{i.e.}, gate-based quantum computing and Ising/annealing model. The first model is discrete and uses programmable (reversible) logic gates acting on qubits in a similar fashion to classical digital architectures. By interconnecting basic logic gates, various quantum algorithms can be implemented that provide computation speed up in comparison to classical counterparts, The work in \cite{BOT} is an informative overview of the application of gate-based quantum algorithms in wireless communication systems, however, gate-based quantum devices are very sensitive to quantum decoherence effects and thus the number of qubits and logic gates that can be applied is limited. The second quantum model (Ising solver) is analogue and relies on the adiabatic principle of quantum mechanics (Adiabatic theorem) \cite{KIM, YAR}. It is mainly used to solve NP-hard combinatorial optimization problems which are modelled as Ising model instances. By controlling the adiabatic time evolution, the system evolves to a final Hamiltonian whose ground state (lowest energy) encodes the solution of the desired (optimization) problem. This is known as quantum annealing and is one of many algorithms/systems that are used for optimization of Ising model instances. 

Ising machines refer to various heuristic solvers designed to find the ground state of the Ising optimization problem (\textit{e.g.}, quantum annealing~\cite{KIM}, coherent Ising machine (CIM)~\cite{HAR}, Oscillator based Ising machine~\cite{oim} \textit{etc}). While the physical implementation of these systems can vary drastically, ranging from interactions between qubits or optical pulses to coupling between oscillators, all Ising machines take an Ising problem as an input and output a candidate solution. These solvers have been mainly used in the communication literature to solve the maximum likelihood (ML) detection problem for large multi-user multiple-input multiple-output (MU-MIMO) setups. The work in \cite{JAM} introduces the QuAMax MU-MIMO detector leveraging tools from quantum annealing. In \cite{JAM2}, the authors exploit parallel tampering (ParaMax) to improve the performance of QuAMax. To overcome error floor effects associated with the straightforward mapping of the ML MU-MIMO decoding problem into the Ising model, the work in \cite{ABH2} adopts the CIM solver that uses an artificial optical spin network to find the ground state of the Ising problem; the proposed CIM-based regularized MU-MIMO detector significantly outperforms previous solutions. A more sophisticated CIM technique is proposed in \cite{ABH1}, which converts the original ML MU-MIMO problem into a perturbation correction problem; this technique provides significant performance gains for high order modulations. The application of Ising machines to complex/combinatorial wireless communication problems seems a promising research direction. 

In this work, we focus on reconfigurable antenna arrays which are an enabling technology for the upcoming 6G communication systems. Reconfigurable antennas have the capability to change in a programmable/controllable way their physical and electrical properties (\textit{i.e.}, configuration) to achieve various objectives (\textit{e.g.}, increase data rate, control interference, boost signal-to-noise ration (SNR) \textit{etc.}). Although the concept is not new \cite{SMI}, recently it has received a lot of attention due to the recent advances in fluid-antenna systems (FAS) \cite{WON}. In FAS, the radiated element of the antenna is liquid which is moving in a controllable way inside a holder; a re-configuration in this case refers to modify the physical position of the liquid. Most of the work in this area focuses on single antenna setups and studies appropriate channel models and/or signal processing techniques that exploit the liquid dimension \cite{KHA, KRI}. The extension of FAS to MIMO settings is an open problem in the literature; the work in \cite{WON2} adopts an information theoretic standpoint of MIMO-FAS while the associated configuration selection problem is overlooked. 

In this paper, we employ antenna configuration selection to maximize SNR for a reconfigurable antenna MIMO system by using tools from quantum computing. More specifically, we adopt the CIM solver which is represented and emulated by a set of stochastic differential equations. The physical problem is firstly formulated as a combinatorial optimization problem with binary variables and multiple constraints (an NP-hard problem). Then, a rigorous mathematical framework that converts the combinatorial problem into an unconstrained quadratic form compatible with CIM implementations is investigated. Numerical results show that the proposed CIM algorithm outperforms classical counterparts and achieves near-optimal performance (similar to exhaustive searching) through appropriate parameterization, while ensuring polynomial complexity. Although we focus on CIMs in this paper, our proposed methods are compatible with any Ising machine including Quantum Annealers. To the best of the authors' knowledge, this is the first time in the literature that CIM tools are used for MIMO configuration selection. 

\noindent {\it Notation:} Lower and upper case bold symbols denote vectors and matrices, respectively, the superscripts $(\cdot)^{T}$ and $(\cdot)^{H}$ denote transpose and conjugate transpose, respectively; $\mathbf{0}$ and $\mathbbm{1}$ denote a vector/matrix with appropriate dimension with all elements equal to zero and one, respectively; $\mathcal{CN}(\mu,\sigma^2)$ represents the complex Gaussian distribution with mean $\mu$ and variance $\sigma^2$; $\textrm{Tr}(\cdot)$ is the trace operator, $\mod$ denotes the modulo operator, and $\lceil \cdot \rceil$ rounds up toward positive infinity.

\section{System model}
We consider a fundamental point-to-point MIMO setup consisting on $N_T$ and $N_R$ antennas at the transmitter and the receiver, respectively. Each antenna is reconfigurable and can change its physical and electrical properties in a controllable way; $N$ distinct states/configurations are assumed. A potential implementation of this setup refers to FAS with $N$ predefined ports at each antenna \cite{WON}; however, the model considered is generic and holds for any type of reconfigurable antenna MIMO. 

\begin{figure}
\includegraphics[width=\linewidth]{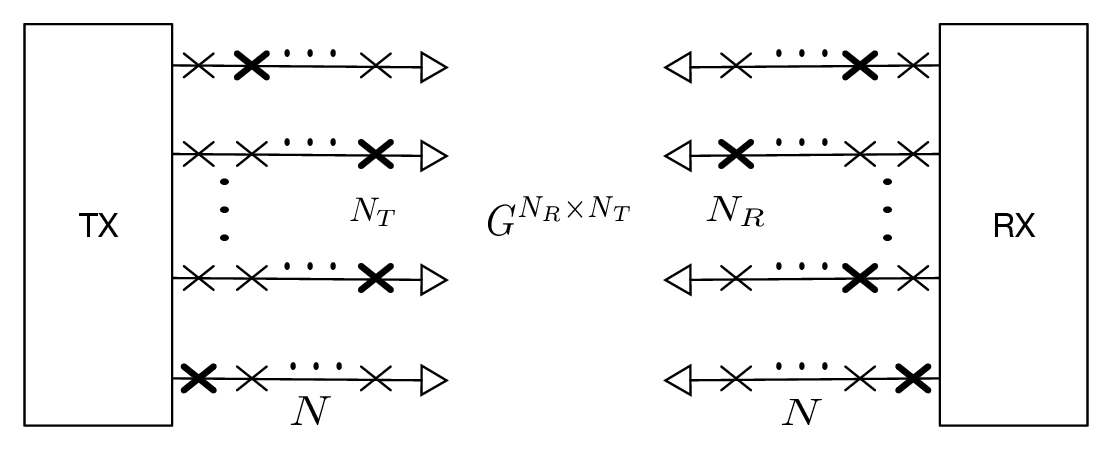}
\vspace{-0.8cm}
\caption{Point-to-point MIMO with $N_T$ and $N_R$ antennas at the transmitter and the receiver, respectively; $N$ configurations in each antenna. The symbol ($\times$) represents an antenna configuration, while the bold symbol ($\boldsymbol{\times}$) corresponds to the selected configuration.}\label{system}
\end{figure}

To facilitate the mathematical formulation, we introduce the complete MIMO channel matrix $\mathbf{G}^{NN_R\times N N_T}$, where entries correspond to the channels between the transmit and and receive antennas for all possible configurations \textit{i.e.}, $g_{i,j}$ is the channel coefficient between the transmit antenna $\lceil j/N \rceil$ with configuration $(j\mod N)$ and the receive antenna $\lceil i/N \rceil$ with configuration $(i\mod N)$. Without loss of generality, we assume independent configurations with Rayleigh block fading channels \textit{i.e.}, $g_{i,j}\sim \mathcal{CN}(0,1)$ \cite{SMI}. 
According to the principles of the reconfigurable antennas, only one configuration/state per antenna can be active in each operation time; the configuration selection reduces the complete channel matrix into the conventional $N_R \times N_T$ MIMO matrix. Fig. \ref{system} schematically depicts the system model.  

The diagonal matrices $\mathbf{X}^{N N_T\times N N_T}$ and $\mathbf{Y}^{N N_R\times N N_R}$ are introduced to enable configuration selection at the transmitter and the receiver side, respectively. The elements of these matrices are binary $x_{i,i},y_{i,i} \in \{0,1\}$ where a value equal to one (or zero) means that the configuration ($i\mod N$) of antenna $\lceil i/N \rceil$ is selected (or not selected). By incorporating the configuration selection into the channel matrix, the equivalent channel becomes equal to $\mathbf{H}=\mathbf{Y}\mathbf{G}\mathbf{X}$. In order to focus our study on the antenna configuration selection process, we assume that configuration selection and power allocation problems are decoupled (similar to \cite{WON2}); for simplicity, a symmetric power allocation\footnote{We note that once the configuration selection problem is solved, the power allocation problem is convex with respect the input covariance matrix and can be solved by using standard convex optimization \cite[Prop. 2.1]{ZHA}.} is considered at the transmitter without loss of generality.   

In this work, reconfigurable antenna MIMO is used to maximize the SNR at the receiver, thus the following combinatorial optimization problem is considered
\begin{subequations}\label{opt11}
\begin{align}
&\max_{\mathbf{X},\mathbf{Y}}\;\textrm{Tr}\{\mathbf{H} \mathbf{H}^H\}\!=\! \textrm{Tr}(\mathbf{X}\mathbf{G}^{H}\mathbf{YYGX})\!=\textrm{Tr}(\mathbf{X}\mathbf{G}^{H}\mathbf{YG}) \label{opt}\\
&\textrm{subject to}\;\sum_{i=kN+1}^{(k+1)N} x_{i,i}=1,\quad k=0,\ldots,N_{T}-1, \label{c1}  \\
&\;\;\;\;\;\;\;\;\;\;\;\;\;\;\;\; \sum_{i=k N+1}^{(k+1)N} y_{i,i}=1,\quad k=0,\ldots,N_{R}-1. \label{c2}
\end{align}
\end{subequations}
The constraints in \eqref{c1} and \eqref{c2} correspond to the physical limitation that only one state is active at each antenna for both the transmitter and the receiver, respectively. It is worth noting that the transmit power, and noise variance are constant terms in the SNR expression and do not affect the optimization problem; they are omitted from the objective function for the sake of simplicity. The considered optimization problem has binary variables and is of combinatorial nature; it is an NP-hard problem with exponential complexity and its optimal solution mainly requires an exhaustive searching over all configurations.     
\subsection{Conventional solutions and benchmarks}
For the formulated problem, we consider three conventional techniques that are used as performance benchmarks for the proposed CIM design. 

\subsubsection{Exhaustive searching (ES)} The ES is the optimal solution and computes the SNR metric for all possible combinations of the antenna configurations. The algorithm requires $N^{N_T} \times N^{N_R}=N^{N_T+N_R}$ SNR calculations (one for each combination) and therefore its complexity becomes exponential with the number of antennas/configurations; it is prohibited for MIMO setups with high number of antennas/configurations.  
\subsubsection{Norm-based selection algorithm (NSA) \cite{SMI}} In the NSA, the receiver and the transmitter select the configurations corresponding to the highest/strongest row and column norms (Euclidean), respectively. More specifically, the selection is firstly performed at one end of the link (\textit{e.g.}, the receiver) and each antenna selects the configuration with the highest row norm; by using the selected row configurations, then each antenna at the other end of the link (\textit{e.g.}, the transmitter) selects the configuration with the highest column norm. This selection scheme reduces significantly the number of calculations \textit{i.e.}, $NN_T+NN_R=N(N_T+N_R)$ and thus it has a high practical interest. 
\subsubsection{Random selection (RS)} The RS is a simple scheme where a random configuration is selected at each antenna; it does not require complicated computations or any intelligence. 

\section{CIM for MIMO configuration selection}

In this section, we firstly highlight the basic properties of the CIM as well as the associated system of stochastic differential equations. Then, the transformation of the considered optimization problem into CIM compatible form is presented.

\subsection{CIM background}
A CIM is a heuristic solver for finding the ground state of an Ising optimization problem, which is a quadratic binary optimization problem and can be expressed as
\begin{equation}
     \arg \min_{\forall i,s_i\in \{-1,1\}}-\sum_{i \neq j}J_{ij}s_{i}s_{j},
\end{equation}
or equivalently in a vector form given by
\begin{equation}
    \arg \min_{\mathbf{s}\in \{-1,1\}^N}  - \mathbf{s}^T\mathbf{J}\mathbf{s}.
\end{equation}.

CIMs were designed to utilize an artificial optical spin network~\cite{marandi2014network} to solve the Ising optimization problems. The dynamics of such systems can be approximately modeled as~\cite{ahc}
\begin{equation}
    \forall i\text{,      }\dfrac{dx_i}{dt} = (p-1)x_i - x_i^3 + \epsilon \sum_{j\neq i} J_{ij}x_j, 
\end{equation}
where $\epsilon=\gamma t$, $\gamma$ and $p$ are model parameters (constant), $x_i$ are state variables describing such systems, and $J_{ij}$ are the coefficients of the Ising problem being solved. While CIMs are designed to find the global optimal, they can get stuck in local minima and limit cycles~\cite{ahc}. An enhanced model with amplitude heterogeneity correction (AHC)~\cite{ahc} that destabilizes these local minima can be used to improve the overall performance. AHC-based CIM model can be described as~\cite{ahc}
\begin{equation}
     \forall i \text{      ,}\dfrac{dx_i}{dt} = (p-1)x_i - x_i^3 + \epsilon e_i \sum_{j\neq i} J_{i,j}x_j,
     \label{eq:cim1}
\end{equation}
\begin{equation}
     \forall i \text{,      } \dfrac{de_i}{dt} = -\beta(x_i^2 - a)e_i,\text{    }e_i > 0,
     \label{eq:cim2}
\end{equation}
where $e_i$, $x_i$ are the state variables of the system, $\beta$ and $a$ are model parameters (constant). The spin solution corresponding to a CIM state is simply given by $s_i = sign(x_i)$, where $sign(\cdot)$ denotes the sign function. In this work, we simulate an AHC-based CIM model by performing numerical integration of (\ref{eq:cim1}) and (\ref{eq:cim2}) for $1000$ time-steps with $dt = 0.01$. We set the parameters of the model as $p=0.98$, $\beta = 1$, $a=2$ and $\epsilon = \gamma t$ with $\gamma = 100$~\cite{ABH1}.

\subsection{CIM formulation: Design}

We present a mathematical framework that transforms the original combinatorial optimization problem in \eqref{opt11} to the Ising model and specifically in an appropriate form that is compatible with  CIM optical hardware implementations. 

Firstly, we introduce a binary vector $\mathbf{b}$ that integrates the diagonal elements from both $\mathbf{X}$, $\mathbf{Y}$ matrices \textit{i.e.},
\begin{align}
\mathbf{b}^T&=\left[x_{1,1},\ldots,x_{NN_{T},NN_T},y_{1,1},\ldots,y_{NN_{R},NN_R}\right] \nonumber \\
&=\left[b_{1},\ldots,b_{N(N_{T}+N_{R})}\right],\;\;\; b_{i}\in\left\{0,1 \right\}.
\end{align}

We also define the symmetric matrix $\mathbf{Q}$ of dimension $N(N_T+N_R)\times N(N_T+N_R)$ given by 
\begin{align}
\mathbf{Q}=\left[ \begin{matrix}
\mathbf{0}_{NN_T\times NN_T} & \frac{1}{2}\mathbf{T} \\[0.5ex] 
\frac{1}{2}\mathbf{T}^T & \mathbf{0}_{NN_R\times NN_R}\\[0.5ex]
\end{matrix}\right],
\end{align} 
\noindent where the matrix $\mathbf{T}^{NN_T\times N N_R}$ consists of the squared amplitudes of the channel coefficients \textit{i.e.},
\begin{align}
\mathbf{T}=\left[ \begin{matrix}
\left| g_{1,1}\right| ^{2} & \left| g_{2,1}\right| ^{2} & \cdots & \left| g_{NN_{R},1}\right| ^{2}\\
\vdots & \vdots & \ddots & \vdots\\
\left| g_{1,NN_{T}}\right| ^{2} & \left| g_{2,NN_{T}}\right| ^{2} & \cdots & \left| g_{NN_{R},NN_{T}}\right| ^{2}\\[0.5ex] 
\end{matrix}\right].
\end{align}
\begin{figure*}[h!]\vspace{-5mm}		
\begin{align}
\mathbf{A}_{k}=\left[\begin{array}{ccc}
				\mathbf{0}_{kN\times kN} & \mathbf{0}_{kN\times N} & \mathbf{0}_{kN\times (N_{T}+N_R-k-1)N}\\
				\mathbf{0}_{N\times kN} & \mathbbm{1}_{N\times N} & \mathbf{0}_{N\times (N_{T}+N_R-k-1)N}\\
				\mathbf{0}_{(N_{T}+N_R-k-1)N\times kN} & \mathbf{0}_{(N_{T}+N_R-k-1)N\times N} & \mathbf{0}_{(N_{T}+N_R-k-1)N\times (N_{T}+N_R-k-1)N}\\
				\end{array} \right] \label{Ak}
\end{align}
	\noindent\makebox[\linewidth]{\rule{18.3cm}{.4pt}}\vspace{-5mm}
\end{figure*}
By using the above definitions, the original optimization problem in \eqref{opt11} can be converted to the following quadratic form:  
\begin{subequations}\label{opt22}
\begin{align}
&\;\;\;\;\;\;\;\;\;\;\;\;\;\;\;\;\;\;\;\;\;\;\;\;\;\;\;\;\;\;\;\;\arg\max_{\mathbf{b}}\;\mathbf{b}^T\mathbf{Q}\mathbf{b} \\
&\textrm{s. t}\;\; P_k(\mathbf{b})=\sum_{i=k N+1}^{(k+1)N}b_i-1=0,\;\;k=0,\ldots,N_{T}+N_{R}-1. \label{opt2}
\end{align}
\end{subequations}

The above quadratic optimization problem has constraints on the binary variables and therefore can not be used directly for quantum implementations. To overcome this bottleneck, we consider the following equivalent quadratic representation of the $k^{th}$ constraint 
\begin{align}
		P_k(\mathbf{b})=\left( \sum_{i=kN+1}^{(k+1)N}b_i-1\right) ^{2}=\left( \sum_{i}b_i\right) ^{2}-2\sum_{i}b_i+1. \label{k-th}
\end{align}

By ignoring the constant terms in \eqref{k-th}, the $k^{th}$ constraint can be written in quadratic binary form as follows
\begin{align}
P_k(\mathbf{b})=\mathbf{b}^{T}\mathbf{A}_{k}\mathbf{b}-2\mathbf{h}_k^{T}\mathbf{b}, 
\end{align}
where $\mathbf{h}_k^{T}=\left[\mathbf{0}_{kN} \quad \mathbbm{1}_{N} \quad \mathbf{0}_{(N_{T}+N_R-k-1)N} \right]^T$ and $\mathbf{A}_k$ is defined in \eqref{Ak}.

Since all the constraints have the same impact on the problem considered, we combine all the constraints in a single aggregate constraint through summation. More specifically, we have 
\begin{align}
P_0(\mathbf{b})&=\sum_{k=0}^{N_T+N_R-1}P_k(\mathbf{b}) \nonumber \\
&=\mathbf{b}^T \left(\sum_{k=0}^{N_T+N_R-1} \mathbf{A}_k \right)-2 \left(\sum_{k=0}^{N_T+N_R-1}\mathbf{h}_k^T \right)\mathbf{b} \nonumber\\
&=\mathbf{b}^T \mathbf{R}\mathbf{b}-2\mathbbm{1}_{N(N_T+N_R)}^T \mathbf{b}. \label{quad_const}
\end{align} 

The next step of the mathematical framework is to convert the binary vector $\mathbf{b}$ into the spin vector $\mathbf{s}$, where the spin variables $s_i$ take values in $\{-1,+1\}$. By considering the transformation $b_i=\frac{1}{2}(s_i+1)$ \cite{KIM}, the expression in \eqref{quad_const} is equivalent to 
\begin{align}
P_0(\mathbf{s})&=\frac{1}{4}\mathbf{s}^T \mathbf{R}\mathbf{s}+\left( \frac{1}{2}\mathbbm{1}_{N(N_T+N_R)}^T \mathbf{R}-\mathbbm{1}_{N(N_T+N_R)}^T\right)\mathbf{s} \nonumber \\
&=\frac{1}{4}\mathbf{s}^T \mathbf{R}\mathbf{s}+\mathbf{q}^T \mathbf{s}, \label{ising}
\end{align}
where the constant terms have been removed since have not any effect on the optimization problem considered. 
The Ising formulation in \eqref{ising} has both linear and quadratic terms, which is not compatible with the CIM optical hardware implementations; we note that the considered CIM architecture requires only quadratic terms \cite{ABH1}. We introduce an auxiliary spin variable $s_\alpha$ to convert the linear terms in \eqref{ising} to quadratic terms~\cite{ABH2}. Specifically, by introducing the extended vector $\mathbf{s}_0^T=[s_{\alpha}\; \mathbf{s}^T]$, the constraint in \eqref{ising} can be written as
\begin{align}
P_0(\mathbf{s}_0)&=\frac{1}{4}\mathbf{s}_0^T\mathbf{R}_0\mathbf{s}_0+\mathbf{s}_0^T
\mathbf{C} \mathbf{s}_0=\mathbf{s}_0^T \left(\frac{1}{4}\mathbf{R}_0+\mathbf{C}   \right)\mathbf{s}_0 \nonumber \\
&=\mathbf{s}_0^T \mathbf{J}_0\mathbf{s}_0, \label{const_f}
\end{align}
where the symmetric matrices $\mathbf{R}_0$ and $\mathbf{C}$ are defined as follows
\begin{equation}
\mathbf{R}_{0}=g(\mathbf{R})=\left[ \begin{array}{c|c}
0  & \begin{matrix} 0 & \cdots & 0 \end{matrix} \\ \hline
\begin{matrix} 0 \\ \vdots \\ 0 \end{matrix} & \mathbf{R}
\end{array}\right],
\end{equation}
\begin{equation}
\mathbf{C}=f(\mathbf{q})=\left[
\begin{array}{c}
\begin{matrix}
0 & \frac{1}{2}\mathbf{q}\\				
\frac{1}{2}\mathbf{q}^T & \mathbf{0}^{N(N_{T}+N_{R})\times N(N_{T}+N_{R})}
\end{matrix} \\
\end{array}\right].
\end{equation}
The last mathematical step for the calibration of the constraint in \eqref{const_f} is to set all the diagonal elements of the matrix $\mathbf{J}_0$ to zero; and then normalize the resulting matrix such as all its entries take values in the range $[-1,+1]$. More specifically, these operations can be represented by the transformation $F_n(\mathbf{J}_0)=\zd(\mathbf{J}_0)/\|\vect(\zd(\mathbf{J}_0))\|_{\infty}$, where $\zd(\cdot)$ sets the diagonal elements to zero, $\vect(\cdot)$ converts a matrix into a vector, and $\| \cdot \|_{\infty}$ denotes the $\infty$- norm ($\max$-norm). Therefore, the constraint in \eqref{const_f} is converted into the CIM- compatible form 
\begin{align}
P_0(\mathbf{s}_0)=\mathbf{s}_0^T F_n(\mathbf{J}_0) \mathbf{s}_0. \label{agg}
\end{align}
By using similar analytical steps, we convert also the quadratic binary objective function in \eqref{opt22} into spin quadratic form without linear terms and normalized quadratic matrix. Specifically, the objective function can be converted as follows 
\begin{align}
\mathbf{b}^T \mathbf{Q} \mathbf{b}\rightarrow \mathbf{s}_0^T F_n(\mathbf{J})\mathbf{s}_0, \label{objf}
\end{align} 
where $\mathbf{J}=\frac{1}{4}g(\mathbf{Q})+f \left(\frac{1}{2}\mathbbm{1}_{N(N_T+N_R)}^T\mathbf{Q} \right)$.

Since CIMs can not handle constraints directly, the last step of the mathematical framework is to combine the objective function in \eqref{objf} with the aggregate constraint in \eqref{agg} by using a penalty scalar $\lambda \in [0,\; 1]$ to ensure the validity of the constraints \cite{YAR}. Since both the objective function and the aggregate constraint are quadratic and normalized, the considered optimization (maximization) problem can take the following final quadratic unconstrained CIM form \textit{i.e.},
\begin{align}
(\overline{s}_{\alpha},\overline{\mathbf{s}})=\arg \max_{s_{\alpha}, \mathbf{s}}\;\;&(1-\lambda)\mathbf{s}_0^T F_n(\mathbf{J})\mathbf{s}_0-\lambda \mathbf{s}_0^T F_n(\mathbf{J}_0)\mathbf{s}_0 \nonumber \\
&=\mathbf{s}_0^T \Big((1-\lambda)F_n(\mathbf{J})-\lambda F_n(\mathbf{J}_0)  \Big) \mathbf{s}_0.
\end{align}

The above (auxiliary) formulation can be solved in the considered CIM architecture and then the produced solution can be used to solve the initial formulation by using the equation $\mathbf{\hat{s}}=\overline{s}_{\alpha}\times \overline{\mathbf{s}}$ \cite{ABH1}. It is worth noting that if the CIM solver does not result in any feasible solution, a random selection algorithm is applied without loss of generality. 

It is obvious that the penalty parameter $\lambda$ is critical for the performance of the CIM algorithm; a larger $\lambda$ enforces feasibility (satisfaction of the constraints) but on the other side less resolution in the objective function and vice-versa. In our numerical studies, this parameter is adjusted empirically through experimentation.  
\begin{figure}
    \includegraphics[width= \linewidth]{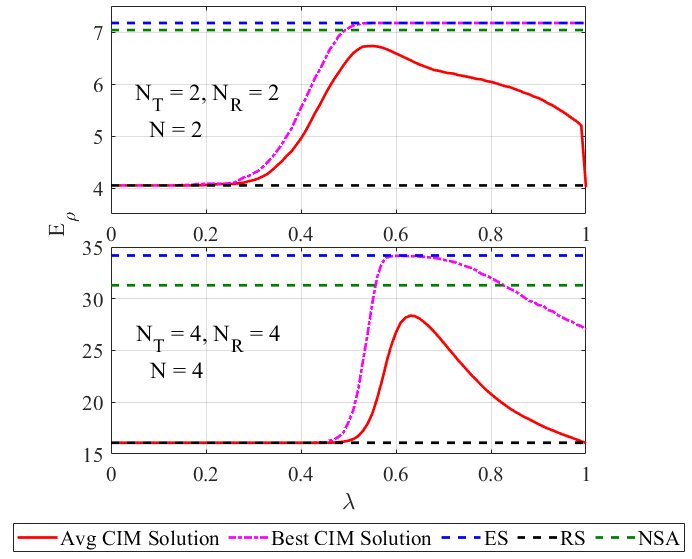}
    \caption{Performance of CIM-based antenna selection for different values of penalty parameter $\lambda$.}
    \label{fig:snr_v_lambda}
\end{figure}
\\
{\it Complexity:} Our proposed formulation uses one spin variable each for representing the selection decision corresponding to $N$ configurations for each $N_T + N_R$ antennas, leading to a total of $N(N_T + N_R)$ spin variables for representing the problem. One auxiliary spin variable is used to convert all the linear terms into quadratic terms~\cite{ABH2}. It takes O($N^2(N_T + N_R)^2$) operations to compute the Ising coefficients corresponding to the SNR maximization objective function ($\mathbf{J}$) and O($N^2(N_T + N_R)^3$) operations to compute Ising coefficients corresponding to the constraint satisfaction ($\mathbf{J}_0$), leading to a total complexity of O($N^2(N_T + N_R)^3$) for computation of the Ising formulation.

\vspace{-0.3cm}
\section{Evaluation}
In this section, we evaluate the performance of our method and benchmark it against ES, NSA and RS schemes. The CIM solver is emulated (in Matlab) by using the classical dynamical system description in \eqref{eq:cim1}- \eqref{eq:cim2}. We focus on two key evaluation metrics:
\begin{itemize}
    \item $E_\rho \triangleq$ Expected value of SNR-maximization objective function over $1,000$ different channel instances (independent Rayleigh fading channels).
    \item $P_c \triangleq$ average probability (over $1,000$ independent Rayleigh fading channels) that CIM generated solution satisfies the problem constraint. 
\end{itemize}
As noted before, CIMs can get stuck in local minima and therefore, it is a common practice to solve the same problem instance multiple times \footnote{It is worth noting that the approach of multiple anneals is common to all QA solvers (\textit{e.g.}, D-WAVE) \cite{JAM,JAM2}; future quantum implementations will further squeeze the anneal time and pro-processing time making QA suitable for real-time applications \cite{KIM}.} , and each of these runs is referred to as an \textit{anneal}~\cite{ABH2}; the best solution among all the anneals is the final output of the CIM algorithm. In this work, we use $1,000$ anneals ($N_a = 1,000$) per problem instance and evaluate both the average performance across all anneals, as well as the performance of the best solution found by the CIM model.
\begin{figure}
    \includegraphics[width= \linewidth]{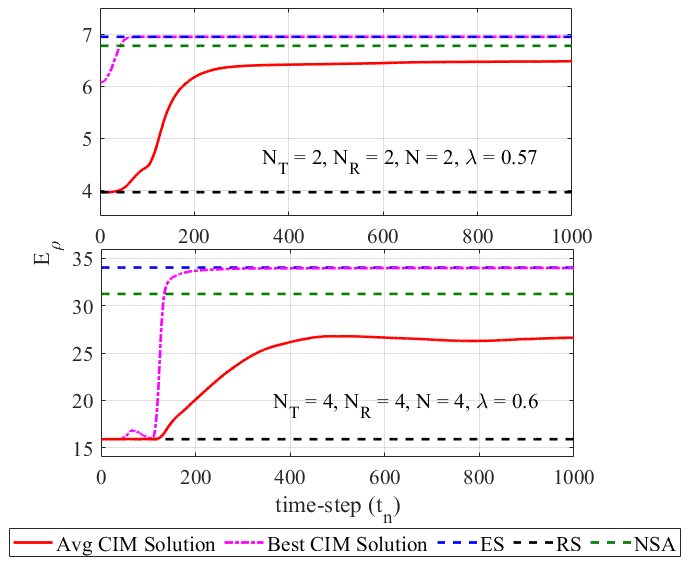}
    \caption{Performance of CIM-based antenna selection as CIM dynamics evolves with time.}
    \label{fig:snr_v_time}
\end{figure}
\begin{figure}
\centering
    \includegraphics[width= 0.9\linewidth]{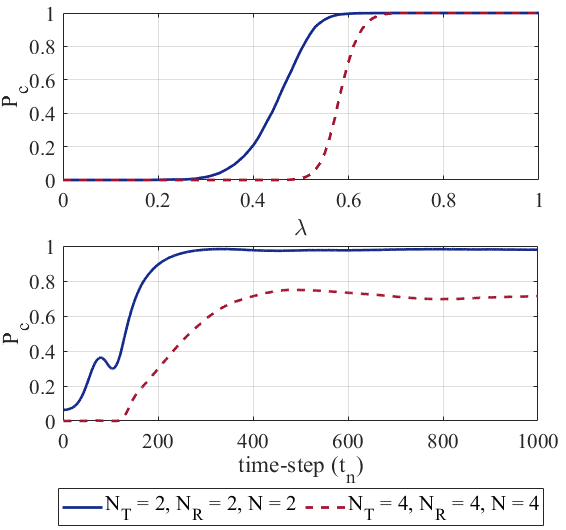}
    \caption{(Top) Probability of constraint satisfaction by CIM-based antenna selection solutions for different values of penalty parameter $\lambda$. (Bottom) Probability of constraint satisfaction by CIM-based antenna selection solutions as CIM dynamics evolve with time.}
    \label{fig:prob_v_lambda}
\end{figure}
\subsection{Varying the penalty parameter $\lambda$}
We vary the penalty parameter $\lambda$ and observe the performance of our design. Recall that $\lambda$ describes the relative weight given to the constraints of the antenna selection problem, where $\lambda = 1$ corresponds to selecting a valid antenna configuration while ignoring the objective function, and $\lambda = 0$ corresponds to optimizing the objective function while completely ignoring the constraints. We simulate two MIMO configurations ($N_T = 2, N_R = 2, N = 2$) and ($N_T = 4, N_R = 4, N = 4$) in Fig.~\ref{fig:snr_v_lambda}. We observe that when $\lambda$ is small, the performance of CIM is similar to random selection. This can be explained by the fact that when $\lambda$ is small, the Ising problem is not able to capture the constraints well, and therefore, the CIM returns invalid solutions (Fig~\ref{fig:prob_v_lambda}~(top)), and the algorithm defaults to RS. When $\lambda$ is very high, the constraints satisfaction dominates the SNR maximization, and therefore CIM solutions are valid but perform worse than ES and NSA. However, we note that at intermediate values of $\lambda$, CIM-based antenna selection can outperform both NSA and ES, and it is possible to empirically tune $\lambda$ to get the best performance.

\subsection{Time-evolution of the CIM solutions}
As noted before, we simulate the behavior of the AHC-based CIM~\cite{ahc} by numerical integration of its dynamical equations. In this section, we select the optimal $\lambda$ based on the empirical analysis in Fig.~\ref{fig:snr_v_lambda} and demonstrate how different performance metrics evolve with time (as we simulate the dynamics of the CIM). In Fig~\ref{fig:snr_v_time}, we plot the expectation of the objective function at each step of numerical integration of the CIM dynamics. We note that, as CIM dynamics evolve with time, its internal states represent a much better solution that progressively improves the objective function. A similar observation holds for the probability of constraint satisfaction, we note that (Fig.~\ref{fig:prob_v_lambda} (bottom)) the internal state of CIM becomes increasingly likely to satisfy the constraints of the problem as CIM dynamics evolve with time. We further note that (from Fig.~\ref{fig:snr_v_time} and Fig.~\ref{fig:prob_v_lambda} (bottom)), both $P_c$ and $E_\rho$ appear to reach a steady state and stop varying after approximately $500$ time-steps, indicating that we need to run the CIM only for that duration. For instance, for the scenario $N_T=N_R=4$ with $N=4$, the exhaustive searching requires $4^8=65,536$ computations, while CIM (best solution) achieves the optimal configurations in $500$ steps/anneals.   

\section{Conclusion}

In this paper, we have studied the problem of antenna configuration selection in reconfigurable antenna MIMO systems. By exploiting the time evolution of a CIM heuristic solver, the antenna configuration that maximizes the SNR at the receiver has been studied. A rigorous mathematical framework that converts the initial constrained binary combinatorial problem (NP-hard) into an unconstrained Ising instance compatible with CIM implementations is developed. The proposed CIM design is studied for different parameters and performance metrics and we show that it achieves near-optimal performance with polynomial complexity. Although our design concerns the CIM heuristic solver, the proposed CIM formulation is generic and can be used by any Ising machine (\textit{e.g.}, quantum annealing). An extension of this work is to integrate the power allocation problem and study also other objective functions \textit{e.g.}, Shannon capacity.

\vspace{-0.1cm}
\section*{Acknowledgements}

This work was funded from the European Research Council
(ERC) under the European Union's Horizon 2020 research and innovation
programme (Grant agreement No. 819819). The work of A. K. Singh and K. Jamieson was funded from the National Science Foundation under Grant No. CNS-1824357.

\bibliographystyle{IEEEtran}
\bibliography{IEEEabrv,ref}

\begin{thebibliography}{10}
\providecommand{\url}[1]{#1}
\csname url@samestyle\endcsname
\providecommand{\newblock}{\relax}
\providecommand{\bibinfo}[2]{#2}
\providecommand{\BIBentrySTDinterwordspacing}{\spaceskip=0pt\relax}
\providecommand{\BIBentryALTinterwordstretchfactor}{4}
\providecommand{\BIBentryALTinterwordspacing}{\spaceskip=\fontdimen2\font plus
\BIBentryALTinterwordstretchfactor\fontdimen3\font minus
  \fontdimen4\font\relax}
\providecommand{\BIBforeignlanguage}[2]{{%
\expandafter\ifx\csname l@#1\endcsname\relax
\typeout{** WARNING: IEEEtran.bst: No hyphenation pattern has been}%
\typeout{** loaded for the language `#1'. Using the pattern for}%
\typeout{** the default language instead.}%
\else
\language=\csname l@#1\endcsname
\fi
#2}}
\providecommand{\BIBdecl}{\relax}
\BIBdecl

\bibitem{TAT}
H.~Tataria, M.~Shafi \emph{et~al.}, ``{6G Wireless Systems: Vision,
  Requirements, Challenges, Insights, and Opportunities},'' \emph{{Proc.
  IEEE}}, vol. 109, pp. 1166--1199, March 2021.

\bibitem{itrs}
ITRS. 2015. International Technology Roadmap for Semiconductors 2.0, executive
  report (2015).

\bibitem{KIM}
M.~Kim, S.~Kasi, P.~A. Lott \emph{et~al.}, ``{Heuristic Quantum Optimization
  for 6G Wireless Communications},'' \emph{IEEE Network}, vol.~35, pp. 8--15,
  August 2021.

\bibitem{ABH1}
A.~K. Singh, D.~Venturelli, and K.~Jamieson, ``{Perturbation-based Formulation
  of Maximum Likelihood MIMO Detection for Coherent Ising Machines},'' in
  \emph{{IEEE Global Commun. Conf.}}, 2022, pp. 2523--2528.

\bibitem{ABH2}
A.~K. Singh, K.~Jamieson, P.~L. McMahon, and D.~Venturelli, ``{Ising Machines'
  Dynamics and Regularization for Near-Optimal MIMO Detection},'' \emph{{IEEE
  Trans. Wireless Commun.}}, vol.~21, pp. 11\,080--11\,094, July 2022.

\bibitem{JAM}
M.~Kim, D.~Venturelli, and K.~Jamieson, ``{Leveraging Quantum Annealing for
  Large MIMO Processing in Centralized Radio Access Networks},'' in
  \emph{Proceedings of the ACM Special Interest Group on Data Communication},
  New York, NY, USA, 2019.

\bibitem{JAM2}
M.~Kim, S.~Mandr\`{a}, D.~Venturelli, and K.~Jamieson, ``{Physics-Inspired
  Heuristics for Soft MIMO Detection in 5G New Radio and Beyond},'' in
  \emph{Proc. Annual Int. Conf. Mobile Comp. Net.}\hskip 1em plus 0.5em minus
  0.4em\relax New York, USA: Association for Computing Machinery, 2021.

\bibitem{BOT}
P.~Botsinis, D.~Alanis, Z.~Babar, H.~V. Nguyen, D.~Chandra, S.~X. Ng, and
  L.~Hanzo, ``{Quantum Search Algorithms for Wireless Communications},''
  \emph{IEEE Commun. Surv. Tutor.}, vol.~21, pp. 1209--1242, 2019.

\bibitem{YAR}
S.~Yarkoni, E.~Raponi \emph{et~al.}, ``{Quantum annealing for industry
  applications: introduction and review},'' \emph{Reports on Progress in
  Physics}, vol.~85, no.~10, p. 104001, Sep. 2022.

\bibitem{HAR}
Y.~Haribara, S.~Utsunomiya, and Y.~Yamamoto, ``{Computational Principle and
  Performance Evaluation of Coherent Ising Machine Based on Degenerate Optical
  Parametric Oscillator Network},'' \emph{Entropy}, 2016.

\bibitem{oim}
T.~Wang and J.~Roychowdhury, ``{OIM: Oscillator-Based Ising Machines for
  Solving Combinatorial Optimisation Problems},'' in \emph{Unconventional
  Computation and Natural Computation}, I.~McQuillan and S.~Seki, Eds.\hskip
  1em plus 0.5em minus 0.4em\relax Cham: Springer Int. Pub., 2019, pp.
  232--256.

\bibitem{SMI}
P.~J. Smith, A.~Firag, P.~A. Martin, and R.~Murch, ``{SNR Performance Analysis
  of Reconfigurable Antennas},'' \emph{IEEE Commun. Letters}, vol.~16, pp.
  498--501, April 2012.

\bibitem{WON}
K.-K. Wong, A.~Shojaeifard, K.-F. Tong, and Y.~Zhang, ``{Fluid Antenna
  Systems},'' \emph{IEEE Trans. Wireless Commun.}, vol.~20, pp. 1950--1962,
  March 2021.

\bibitem{KHA}
M.~Khammassi, A.~Kammoun, and M.-S. Alouini, ``{A New Analytical Approximation
  of the Fluid Antenna System Channel},'' \emph{IEEE Trans. Wireless Commun.},
  vol.~22, pp. 8843--8858, Dec. 2023.

\bibitem{KRI}
C.~Psomas, G.~K. Kraidy, K.~K. Wong, and I.~Krikidis, ``{On the diversity and
  coded modulation design of fluid antenna systems},'' \emph{IEEE Trans.
  Wireless Commun.}, vol.~23, pp. 2082--2096, March 2024.

\bibitem{WON2}
W.~K. New, K.-K. Wong \emph{et~al.}, ``{An information-theoretic
  characterization of MIMO-FAS: optimization, diversity-multiplexing tradeoff
  and q-outage capacity},'' \emph{IEEE Trans. Wireless Commun. (to appear)},
  vol. preprint arxiv: 2303.02269, 2023.

\bibitem{ZHA}
R.~Zhang and C.~K. Ho, ``{MIMO Broadcasting for Simultaneous Wireless
  Information and Power Transfer},'' \emph{IEEE Trans. Wireless Commun.},
  vol.~12, pp. 1989--2001, May 2013.

\bibitem{marandi2014network}
A.~Marandi, Z.~Wang, K.~Takata, R.~L. Byer, and Y.~Yamamoto, ``{Network of
  time-multiplexed optical parametric oscillators as a coherent Ising
  machine},'' \emph{Nature Photonics}, 2014.

\bibitem{ahc}
T.~Leleu, Y.~Yamamoto, P.~L. McMahon, and K.~Aihara, ``{Destabilization of
  Local Minima in Analog Spin Systems by Correction of Amplitude
  Heterogeneity},'' \emph{Phys. Rev. Lett.}, vol. 122, p. 040607, Feb 2019.

\end{thebibliography}

\end{document}